\documentclass[prl,aps,showpacs]{revtex4}
\usepackage{graphicx}
\usepackage[dvips]{color}
\newcommand{\moyal}{\text{Tr } L_q^2}

\begin{document}
\title{Parameter scaling in a novel measure of quantum-classical 
difference for decohering chaotic systems}

\author{Nathan Wiebe$^{(a)}$, Parin Sripakdeevong$^{(b)}$, 
Arnaldo Gammal$^{(c)}$, and Arjendu K. Pattanayak$^{(b)}$}
\date{\today}
\affiliation{
(a) Institute for Quantum Information Science, University of Calgary,
Alberta T2N 1N4, Canada
(b) Department of Physics and Astronomy, Carleton College, 
Northfield, Minnesota 55057
(c)Instituto de F\'isica, Universidade de S\~{a}o Paulo,
05508-090, S\~ao Paulo-Brasil \\
}

\begin{abstract}
In this paper we introduce a diagnostic for measuring the
quantum-classical difference for open quantum systems, which is the
normalized size of the quantum terms in the Master equation for 
Wigner function evolution. 
For a driven Duffing oscillator, this measure shows remarkably precise 
scaling over long time-scales with the parameter $\zeta_0=\hbar^2/D$.
We also see that, independent of $\zeta_0$ the 
dynamics follows a similar pattern. For small $\zeta_0$ all of our 
curves collapses to essentially a single curve when scaled by the 
maximum value of the quantum-classical difference. In both limits of
large and small $\zeta_0$ we 
see a saturation effect in the size of the quantum-classical difference; 
that is, the instantaneous difference between quantum and classical 
evolutions cannot be either too small or too large. 

\end{abstract}
\maketitle

The quantum--classical transition for open quantum systems is important to 
understand for fundamental and practical reasons, including the design 
of quantum computers. For quantum systems with classical analogues permit 
chaos this transition is particularly interesting because these systems 
often display unusual quantum effects, such as rapid entanglement 
generation~\cite{ghose} and hyper-sensitivity to perturbation~\cite{haake},
and it would be useful to understand when such effects appear. It is an 
inherently multi-parameter transition depending on the relative size of 
$\hbar$ compared to the characteristic action and on the strength of the 
interaction between the system and its environment as measured by some 
parameter $D$. The chaotic behavior of the classical limit is also crucial, 
as seen in their decoherence dynamics argued to be related to the Lyapunov 
exponents of the system.  An efficient approach that allows for broader
conclusions is to use~\cite{akp-03} scaling in composite parameters, 
which indicates~\cite{scaling-confirm} that the quantum-classical 
difference as measured by some quantity $QC_d(\hbar,D,\lambda)$ is a 
function $QC_d'(\zeta)$ of a single composite parameter 
$\zeta =\hbar^\alpha D^\beta\lambda^\gamma$. 

A good measure $QC_d$ is intuitive, global, and easy to compute.
A Kullback-Liebler-like distance between classically propagated and 
quantally propagated distributions picks out scaling properties 
cleanly~\cite{akp-03} when computed for a 2-phase-space dimension 
map. Overlaps functions are however, difficult to compute in 
greater dimensions, or for flows. Studying the quantal dynamics
entropy $S_2=\ln(P)$, where $P = {\rm Tr}(\rho^2)$ also yielded 
useful insights. However, the scaling did not last very long in 
that instance, and the direct relationship to quantum-classical
difference is not clear; the use of the entropy has also been otherwise 
critiqued~\cite{wisniacki}. We report here on a measure that is 
(a) time-dependent, (b) explicitly measures the difference between 
quantum and classical evolution, and (c) is calculated from the 
evolution of a single distribution.  This measure shows remarkable 
scaling, in parameter space and over long time-scales,
and we consequently uncover interesting insights into unexpected behavior
in the dynamics. Specifically, in the classical limit, the relative 
size of the quantum and classical terms saturates, so that 
quantum-classical differences continue to be propagated, rather 
than decreasing with time as might be naively expected. Conversely,
at the near-quantal limit of small $D$ and large $\hbar$ the 
quantum terms remain comparable to the classical instead of
dominating. Taken together, this quantifies the smoothness of 
the quantum-classical transition for open systems compared to closed 
systems.

We start by considering the Master equation for a Wigner function $\rho_W$,
or quantum quasi-probability, evolved under Hamiltonian flow with
potential $V(q)$ while coupled to an external environment~\cite{zp}:
\begin{eqnarray}
{\partial \rho_W\over\partial t}
&=& L_c + L_{q} + T\nonumber \\
&=&\{H,\rho_W\} 
+\sum_{n \geq 1}\frac{\hbar^{2n}(-1)^n}{2^{2n} (2n +1)!}
\frac{\partial^{2n+1} V(q)}{\partial q^{2n+1}}\;
\frac{\partial^{2n+1} \rho_W}{\partial p^{2n+1}}
+ D \nabla_p^2\rho_W
\label{wigner}
\end{eqnarray}
The Poisson bracket $L_c$ generates the classical evolution for
$\rho_W$, the quantal $\hbar$ terms are denoted by $L_q$ and the 
environmental coupling $T$ is modeled by a diffusive term with 
coefficient $D$. The 
computational results presented below use coupling only to the momentum 
variables. For analytical simplicity, we assume coupling to all 
phase-space variables, justified since the dynamical chaos 
mixes the various phase-space directions.

\begin{figure}[htbp]
\centering
{\includegraphics[width=0.7\textwidth]{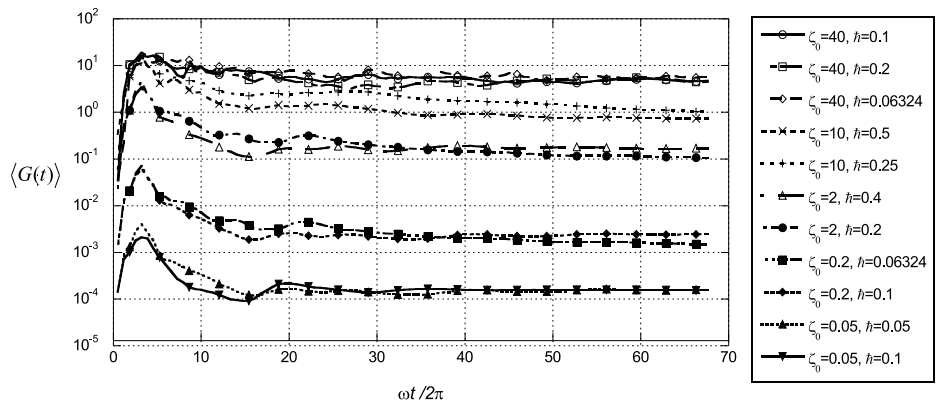}}
\caption {Quantum-classical difference as measured by $\langle G(t)\rangle$,
which is $G(t)$ averaged over one driving period, for
various initial parameters. The $y$-axis is logarithmic. Notice the
remarkable scaling with the composite parameter $\zeta_0 = \hbar^2/D$.
For the initial Gaussian to be well-localized in the chaotic region,
$\sigma^2_q \leq 0.05$. Hence
for our calculations, we set $\sigma^2_q =0.05$ and determine $\sigma^2_p$
from the constraint, $\sigma_q \sigma_p=\hbar/2$, imposed by
minimum-uncertainty condition}
\label{figone}
\end{figure}

The measure we propose in this paper is a normalized average of the 
square of $L_q$ (see Eq.(~\ref{wigner}) for the precise definition). 
The square before the averaging is necessary to compensate for the 
arbitrary sign of $L_q$ in different regions of phase-space. For
example, in the driven Duffing problem that we study below and which is
given by $H = p^2/2m -Bx^2 + (C/2)x^4 + Ax\cos(\omega t)$
the quantity ${\rm Tr}[L_q]$ is identically $0$. The normalization
makes the measure dimensionless, yielding 
\begin{equation}
G(t) = \frac{{\rm Tr}[L^2_q]}{{\rm Tr}[\big(\partial_t\rho_w(t)\big)^2]}.
\end{equation}
Physically $G$ is the relative size of the quantum part of the evolution 
for the Wigner function, and it can exceed unity through partial 
cancelation between $L_q$ and the Poisson bracket in the evolution 
equation for $\rho_w$. Note that $G$ does not measure the classicality 
of a state, but rather the classicality of its evolution and is
analogous to the time-derivative of previous-used measures.

In Fig.~\ref{figone}, we plot $G(t)$ for various combinations of
$\hbar$ and $D$ {for a Duffing oscillator with the parameters
$(m = 1,B = 10,C = 1,A = 1,$ and  $\omega = 5.35)$}. These results
are calculated starting with a minimum uncertainty Gaussian Wigner
function that is well localized in the chaotic region and propagating
it using Eq.~(\ref{wigner}) as has been previously
done\cite{mp,gammal-07}. The figure shows {\em remarkable} scaling 
behavior in the measure $G(t)$. That is, the dynamics of the 
quantum-classical difference are seen to depend {\em only} on the 
composite parameter $ \zeta_0 = \hbar^2/D$ over a factor of $800$ 
in its value.  Although such behavior has been seen earlier for 
short times with the entropy~\cite{akp-03,gammal-07}, in this case 
the scaling dynamics lasts for the full time-scale monitored by us.

The absolute size of $G$ grows with $\zeta_0$, and is much greater for 
$\zeta_0 \simeq 40$ compared to $\zeta_0 \simeq 0.05$, which is
reassuringly physically intuitive. Specifically, the classical limit is
the regime when $\zeta_0\leq 0.2$, where $G$ is always so small that 
the differences between the evolution of the Wigner function $\rho_W$
and its classical counterpart $\rho_C$ are negligible. 

Independent of the value of $\zeta_0$, we see similar dynamical behavior
with identical stages: We start with (i) a rapid increase in $G$, which 
can be understood as essentially the behavior of a closed quantum system 
since the gradients of the distribution have not increased enough for the 
diffusive term to become relevant. This is followed by (ii) a turnover and 
an exponential decrease in $G$ as the distribution starts filling the 
phase-space and the diffusive terms kick in -- the overall distribution 
continues to evolve, but the relative size of the quantal terms is now
decreasing as a result. Finally, (iii) at long times $G$ saturates; 
this happens because the distribution has almost relaxed to its final
state. At this point final stage, both ${\rm Tr}[L_q(t)^2]$ and 
${\rm Tr}[\rho_w(t)^2]$ decrease exponentially with time. That is, at
longer times, the quantal and classical contributions to the evolution 
of the Wigner function reach a steady-state ratio.

\begin{figure}[t!]
\centering
{\includegraphics[width=1\textwidth]{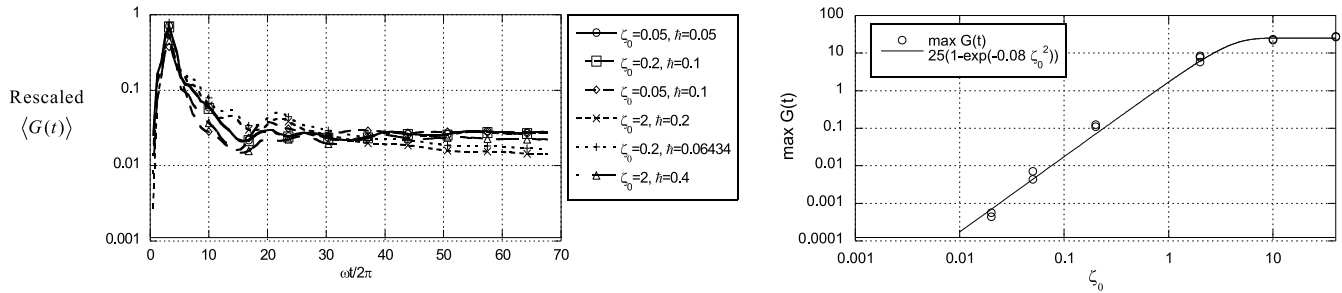}}
\caption{This plot shows in when we re-scale $G(t)$ by dividing the data
for $\zeta\leq2$ in Figure~\ref{figone} by the maximum value of $G(t)$,
the data collapses to approximately a single curve.  Also this data shows
that the maximum value of $G(t)$ as a function
is described very well by
a Gaussian in $\zeta_0$.}
\label{figtwo}
\end{figure}	

To examine the functional dependence of $G(t)$ on $\zeta_0$, we start at
$\zeta_0\rightarrow 0$, the classical limit for open quantum systems.
Fig.~\ref{figtwo} shows that in the classical limit the greatest value
of $G(t)$ for small $\zeta_0$ is approximately \begin{equation}\label{maxg}
\max_{t}G(t)\approx 25\left(1-\exp(-0.08\zeta_0^2) \right).
\end{equation}
That is, the maximum value of $G(t)$ scales approximately quadratically with
$\zeta_0$ in the limit of small $\zeta_0$ and is \emph{only} a function
of $\zeta_0$. This is a remarkably time-independent
relationship, and in Fig.~(\ref{figtwo}) we show that for
$\zeta_0\leq 2$, $G(t)$ collapses to essentially a single function
of $\zeta_0$ when we divide through by the maximum value given in
Eq.~(\ref{maxg}).

\begin{figure}[htbp]
\centering
{\includegraphics[width=0.7\textwidth]{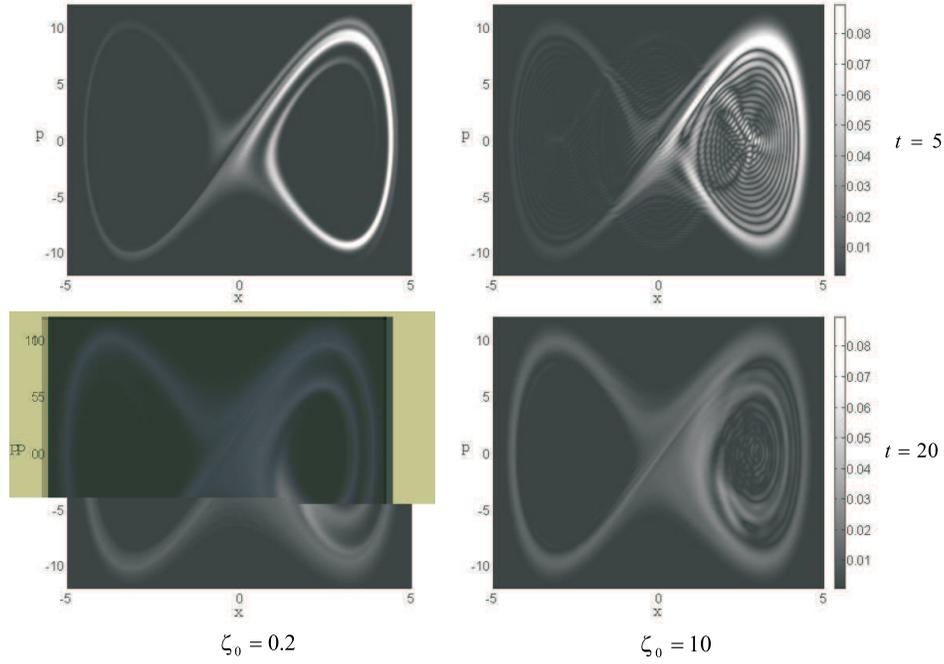}}
\caption{This plot shows the absolute value of $\rho_w$ for
$\zeta_0=0.2$ $(\hbar=0.01, D=5\times 10^{-4})$ and $\zeta_0=10$ 
$(\hbar=0.125, D=1.5625\times 10^{-3})$ evaluated at times ranging from
$t=5$ to $t=20$. The Wigner function for $\zeta_0=10$ shows clear
signs of quantum interference, but the data for $\zeta_0=0.2$
shows no visible signs of quantum behavior at the times considered.
This reflects the observation that $G(t)$ is a slowly decreasing
function for $\zeta_0=10$ because the quantum effects plotted are
also persistent with time.}
\label{figthree}
\end{figure}

The transition out of the near-classical regime starts at 
$\zeta_0\approx 2$. As $\zeta$ increases further, we expect quantum
effects to increase and might naively predict that
$\moyal\approx\text{Tr} (\partial_t
\rho_w(t))^2$ before exceeding it and then becoming the dominant term.
However, the saturation effects in Fig.~(\ref{figone}) indicate
otherwise. Further, plots (Fig.~(\ref{figthree})) of the absolute value of
the Wigner function for the Duffing Oscillator evaluated at the
times $t=5\text{ and } 20$ for $\zeta_0=0.2$ and $\zeta_0=10$,
clearly show that as quantum interference becomes more substantial,
the amplitude of interference fringes approaches that of the
classical phase space structure underlying it. However, these
classical structures (in this case the noise-broadened homoclinic
tangle of the stable and unstable manifolds) always retain a
substantial contribution, again pointing to the wisdom of considering
the classical dynamics when thinking about quantum chaotic systems.

\begin{figure}[htbp]
\centering
{\includegraphics[width=1\textwidth]{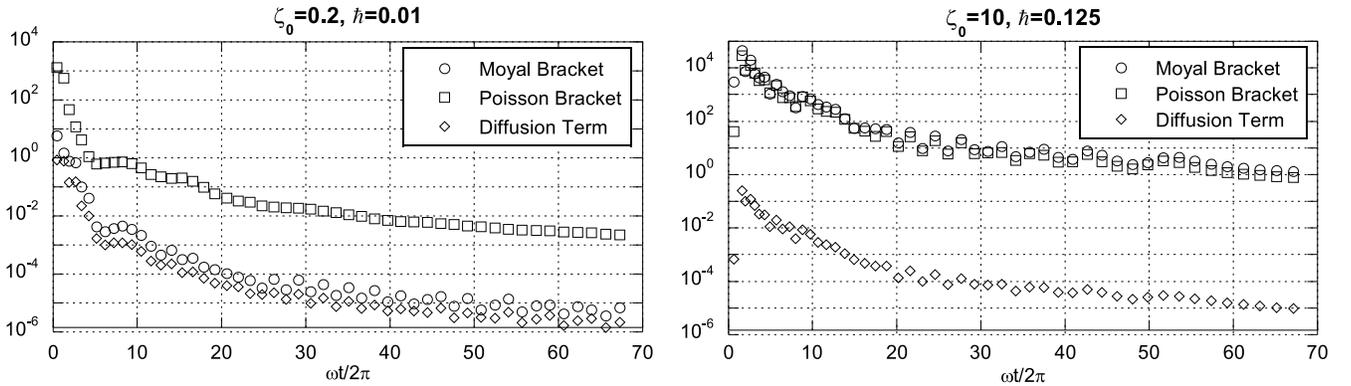}}
\caption{This plot shows that for $\zeta_0=10$ with
$(\hbar=0.125, D=1.5625\times 10^{-3})$ the classical terms
are roughly comparable to the quantal terms. For $\zeta_0=0.2$
with $(\hbar=0.01, D=5\times 10^{-4})$ the classical terms are substantially
larger than the quantum terms. The diffusion term is negligible compared
to the classical in both cases.
}\label{figfour}
\end{figure}		

This saturation effect is examined in greater detail in
Fig.~\ref{figfour}, in which it is shown that the observed saturation 
effect occurs because
the magnitude of $\text{Tr} L_c^2$ becomes comparable to $\text{Tr} 
L_q^2$, while the diffusion term remains negligible. This saturation can be 
understood in the asymptotic regime by considering the influence of 
quantum interference effects on $\rho_w$.

Quantum mechanical effects often appear in physical systems as interference
patterns with characteristic wavenumbers that scale with $\hbar^{-1}$.
Therefore we expect that if $\hbar^2/D$ is large we expect that
$\partial_p \rho_w \sim \hbar^{-1}$ and
$\partial_x\rho_w\sim\partial_p\rho_w \sim\hbar^{-1},
\hbar^2\partial_p^3 \rho_w\sim \hbar^{-1}  \text{ and }
D\partial_p^2\rho_w\sim \frac{D}{\hbar^2}$ for a chaotic system.
Hence the quantal terms should be comparable to the classical in the
limit of large $\zeta_0$ if $\rho_w$ is dominated by interference effects,
which is reasonable for our $\zeta_0=10$ data according to
Fig.~(\ref{figthree}).

To summarize, we considered the normalized size of the quantum terms in the
Master equation for Wigner function evolution in an open quantum system.
For a driven Duffing oscillator, this measure shows remarkably precise 
scaling over long time-scales with the parameter $\zeta_0=\hbar^2/D$.
We also see that, independent of $\zeta_0$ the dynamics follows a
similar pattern. For small $\zeta_0$ all of our curves collapses to 
essentially a single curve when scaled by the maximum value of the
quantum-classical difference. In both limits of large and small $\zeta_0$ 
we see a saturation effect 
in the size of the quantum-classical difference; that is, the instantaneous 
difference between quantum and classical evolutions cannot be either too 
small or too large. This further confirms the growing intuition that 
decoherence softens the quantum-classical transition for nonlinear
systems. Open questions include whether
this remarkable scaling of $G(t)$ with $\zeta_0$ is exhibited for other
Hamiltonians, such as the driven rotor, whose Moyal series does not terminate.
	
{\em Acknowledgments}: N.W. is supported by the MITACS research network. 
A.G. is partially supported by FAPESP (Brazil) and CNPq (Brazil). 
A.K.P. acknowledges a CCSA Award from Research Corporation, and support 
from the SIT, Wallin, and Class of 1949 Funds from Carleton College as 
well as hospitality from CiC (Cuernavaca) during this work.


\begin{thebibliography}{99}

\bibitem{ghose} S.~Ghose, R.~Stock, P.~Jessen, 
L.~Roshan, and A.~Silberfarb, \pra {\bf 78}, 042318 (2008); C.M. Trail,
V. Madhok, and I.H. Deutsch \pre {\bf 78}, 046211 (2008).

\bibitem{haake} F.~Haake, ``Quantum Signatures of Chaos'' 
(Springer-Verlag, Berlin, 1991).

\bibitem{mp} D. Monteoliva and J.P.~Paz, \prl {\bf 85}, 3373 (2000).
	
\bibitem{zp}W.H.~Zurek and J.P.~Paz, \prl {\bf 72}, 2508 (1994);
Physica {\bf 83 D}, 300 (1995).
	
\bibitem{akp-03} A.K. Pattanayak, B.~Sundaram, and B.~D.~Greenbaum,
\prl {\bf 90}, 014103 (2003).
	
\bibitem{scaling-confirm} N.~Wiebe and L.~Ballentine, \pra {\bf 72}, 022109
(2005); F.~Toscano et al., \pra {\bf 71}, 010101 (R) (2005); A.R.R.~Carvalho
et al., Phys. Rev. E 70, 026211 (2004).

\bibitem{gammal-07}A.~Gammal and A.~K.~Pattanayak, \pre {\bf  75}, 036221
(2007).

\bibitem{wisniacki}D.~A.~Wisniacki and F.~Toscano, \pre {\bf 79},
025203(R) (2009).
	
\end{thebibliography}
\end{document}